\documentclass[prb,twocolumn,reprint,superscriptaddress]{revtex4-1}
\usepackage[utf8]{inputenc}
\usepackage{amsmath}
\usepackage{amssymb}
\usepackage{bm}
\usepackage{graphicx}
\usepackage{graphics}
\usepackage{color}
\usepackage{textcomp}
\usepackage{subfigure}
\usepackage[super]{nth}
\def\be{\begin{equation}}
\def\ee{\end{equation}}
\def\ba{\begin{eqnarray}}
\def\ea{\end{eqnarray}}

\def\pet3{Pd$_x$ErTe$_3$}

\begin{document}


\title{Robust superconductivity intertwined with charge density wave and disorder in Pd-intercalated ErTe$_3$}


\author{Alan Fang}
\affiliation{Stanford Institute for Materials and Energy Sciences,\\
SLAC National Accelerator Laboratory, 2575 Sand Hill Road, Menlo Park, CA 94025}
\affiliation{Geballe Laboratory for Advanced Materials, Stanford University, Stanford, CA 94305}
\affiliation{Department of Applied Physics, Stanford University, Stanford, CA 94305}

\author{Anisha G. Singh}
\affiliation{Geballe Laboratory for Advanced Materials, Stanford University, Stanford, CA 94305}
\affiliation{Department of Applied Physics, Stanford University, Stanford, CA 94305}

\author{Joshua A. W. Straquadine}
\affiliation{Geballe Laboratory for Advanced Materials, Stanford University, Stanford, CA 94305}
\affiliation{Department of Applied Physics, Stanford University, Stanford, CA 94305}

\author{Ian R. Fisher}
\affiliation{Stanford Institute for Materials and Energy Sciences,\\
SLAC National Accelerator Laboratory, 2575 Sand Hill Road, Menlo Park, CA 94025}
\affiliation{Geballe Laboratory for Advanced Materials, Stanford University, Stanford, CA 94305}
\affiliation{Department of Applied Physics, Stanford University, Stanford, CA 94305}

\author{Steven A. Kivelson}
\affiliation{Stanford Institute for Materials and Energy Sciences,\\
SLAC National Accelerator Laboratory, 2575 Sand Hill Road, Menlo Park, CA 94025}
\affiliation{Geballe Laboratory for Advanced Materials, Stanford University, Stanford, CA 94305}
\affiliation{Department of Physics, Stanford University, Stanford, CA 94305}

\author{Aharon Kapitulnik}
\affiliation{Stanford Institute for Materials and Energy Sciences,\\
SLAC National Accelerator Laboratory, 2575 Sand Hill Road, Menlo Park, CA 94025}
\affiliation{Geballe Laboratory for Advanced Materials, Stanford University, Stanford, CA 94305}
\affiliation{Department of Applied Physics, Stanford University, Stanford, CA 94305}
\affiliation{Department of Physics, Stanford University, Stanford, CA 94305}


\date{\today}

\begin{abstract}
Pd-intercalated ErTe$_3$ is studied as a model system to explore the effect of  ``intertwined'' superconducting and charge density wave (CDW) orders. Despite the common wisdom that superconductivity emerges only when CDW is suppressed, we present data from STM and AC susceptibility measurements that show no direct competition between CDW order and superconductivity.  Both coexist over most of the intercalation range, with uniform superconductivity over length scales that exceed the superconducting coherence length.  This is despite persisting short-range CDW order and increased scattering from the Pd intercalation. While superconductivity is insensitive to local defects in either of the bi-directional CDWs, vestiges of the Fermi-level distortions are observed in the properties of the superconducting state.
\end{abstract}


\maketitle


\section{Introduction}

Recently the concept of ``intertwined order,'' \cite{FradRMP2015} in which the same features of the microscopic physics produce multiple ordering tendencies with similar energy or temperature scales,  has emerged as key to understanding the complex phase diagrams of strongly correlated electron systems. In particular, the 
relation between superconductivity 
and  finite-range charge density wave (CDW) correlations 
has attracted much attention, 
in part because of its possible relevance to high temperature superconductors such as the cuprates. Furthermore, 
other material systems that exhibit simultaneous CDW order and superconductivity, such as NbSe$_2$ (e.g. \cite{Weber2011,Cho2018}), are now being revisited
in this context, despite the fact that they were previously thought to be well understood.
Adopting  a standard theoretical strategy for such problems, 
we wish to identify a well characterized model system in which CDW and superconductivity mutually exist, but at the same time can be manipulated to alter one or the other forms of order,  
so as to explore their intertwined behavior.

\begin{figure}[h]
\includegraphics[width=1.0\columnwidth]{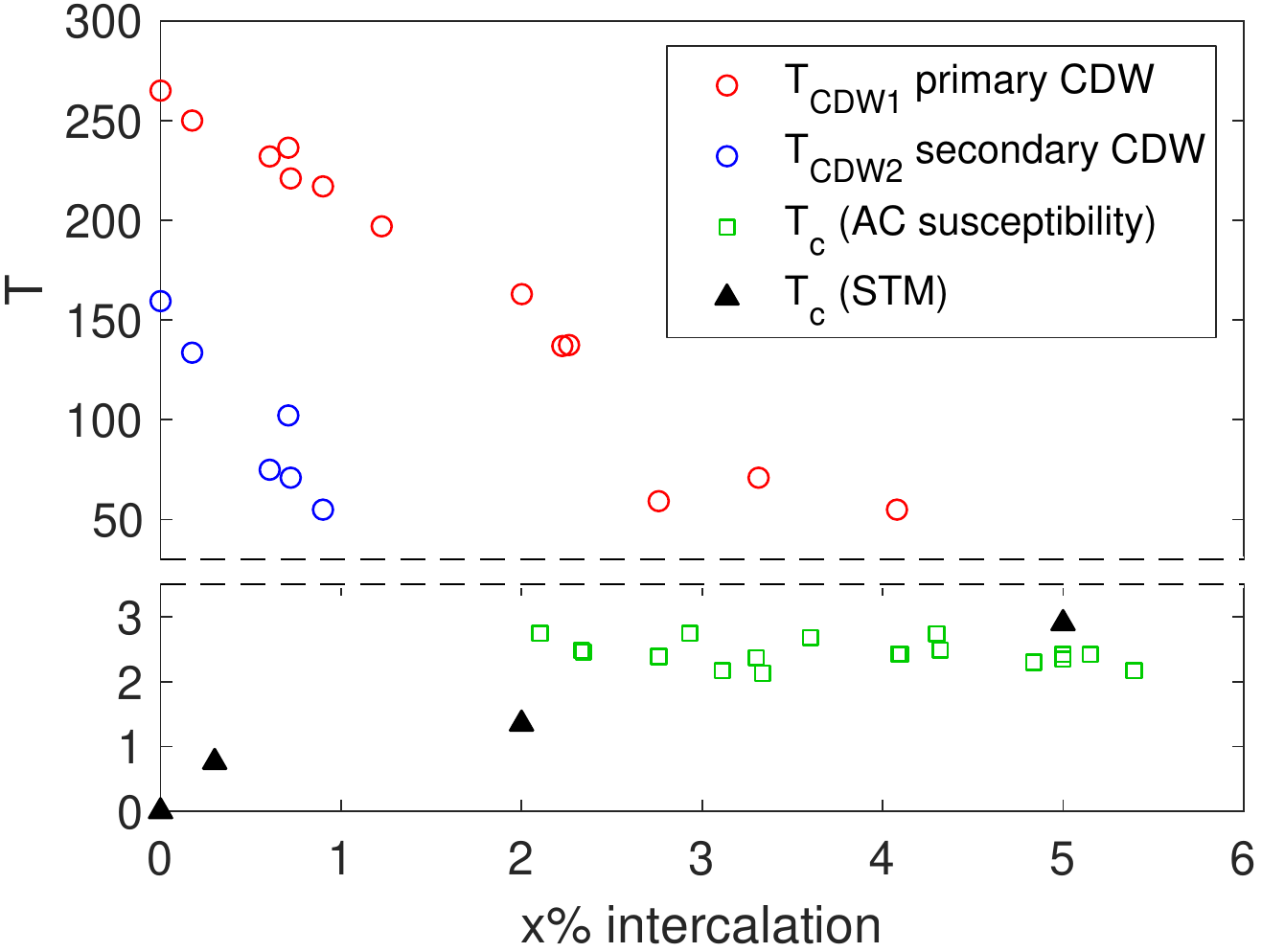}
\caption{ Phase diagram for Pd$_{x}$ErTe$_{3}$ indicating the variation with Pd content ($x$) of (lower panel) the superconducting critical temperature and (upper panel) the vestiges of the two CDW transitions. In the presence of disorder induced by the Pd intercalants, the sharp CDW phase transitions found for the pristine ($x=0$) compound are progressively smeared into cross-overs. The CDW transition temperatures and some of the $T_{c}$ measurements were previously reported in \cite{Straquadine2019}. }
\label{pd}
\end{figure}

 ErTe$_3$ has been shown to exhibit two separate, mutually perpendicular, incommensurate unidirectional CDW phases setting in at $T_{CDW1}=270$ K  for the primary, and $T_{CDW1}=165$ K for the secondary transitions respectively \cite{Moore2010}. It has further been demonstrated that Pd intercalation 
supresses CDW long-range order in  
Pd$_x$RTe$_3$,\cite{Straquadine2019,fang2019disorder} and that this allows  superconductivity to emerge with a transition temperature, $T_c\sim 3$K, that is roughly independent of  $x$ for $x\gtrsim 2$\%. 
(See Fig.~\ref{pd}).
A superficially related observation is that hydrostatic pressure suppresses the CDW order in such a way that it ultimately gives way to a CDW-free superconducting state with an approximately pressure independent $T_c$ of comparable magnitude.
 
 Pd ions intercalate between the van der Waals bonded Te bilayers, leading to local perturbations in the periodic potential. In contrast to the action of uniform hydrostatic pressure, 
the disorder-induced suppression of long range CDW formation, as visualized via STM measurements \cite{fang2019disorder}, occurs via a proliferation of topological defects (dislocations) below the original critical temperature, $T_{CDW}$. 
Indeed, for small but non-zero $x$, the CDW ordered phase gives way to Bragg-glass phases. 
Apparently the presence of dense short-range CDW correlations has little effect on the emergent superconductivity.

In this paper we present a scanning tunneling microscopy and spectroscopy study, complemented by AC susceptibility measurements, to map the evolution of superconducting features with rising intercalation level.  Fig.~\ref{pd} contains a summary of the transition temperature data from resistivity for the CDW transitions \cite{Straquadine2019}, and AC susceptibility and STM for superconducting transitions.  We probe additional spatial properties such as uniformity of gap size and coherence length. 
Specifically, we show that the superconducting order coexists uniformly with the strong short-range CDW order, and that the short-range CDW order is ``gapless'' in the sense that the low energy density of states shows no significant suppression.  The implications of these observations for the mechanism of CDW formation and for the unexpectedly minimal ``competition'' between CDW and superconducting order are discussed.

\section{Experiment}
Pd$_x$ErTe$_3$ samples were grown using a Te self-flux for pure RTe$_3$ compounds \cite{Ru2006}, with the addition of small amounts of Pd to the melt. A detailed description of sample preparation, characterization and the effect of Pd intercalation on the bulk properties are given in \cite{Straquadine2019}.  AC susceptibility measurements were performed in a Quantum Design MPMS3 at frequencies of 75Hz and 757Hz, while a magnetic field was applied both parallel and perpendicular to the $b$-axis of the sample (out of plane), on samples of 2.8\%, 3.3\%, 3.6\%, and 4.1\% Pd intercalation.  Samples were zero-field cooled before a field between 0-500 Oe was applied. 

Scanning tunneling microscopy (STM) and Spectroscopy (STS) were performed with a hybrid UNISOKU USM1300 system plus home-made Ultra High Vacuum (UHV) sample preparation and manipulation system.  The samples were cleaved at pressures of low $10^{-10}$ torr and immediately transferred to the low temperature STM.  Measurements were done at $\approx$ 370-400 mK. Four levels of intercalation were studied in the STM: 0\% (pristine), 0.3\%, 2\% and 5\% (marked in Fig.~\ref{pd}). 

\section{Results}
\subsection{AC Susceptibility}

\begin{figure}[h]
\includegraphics[width=\columnwidth]{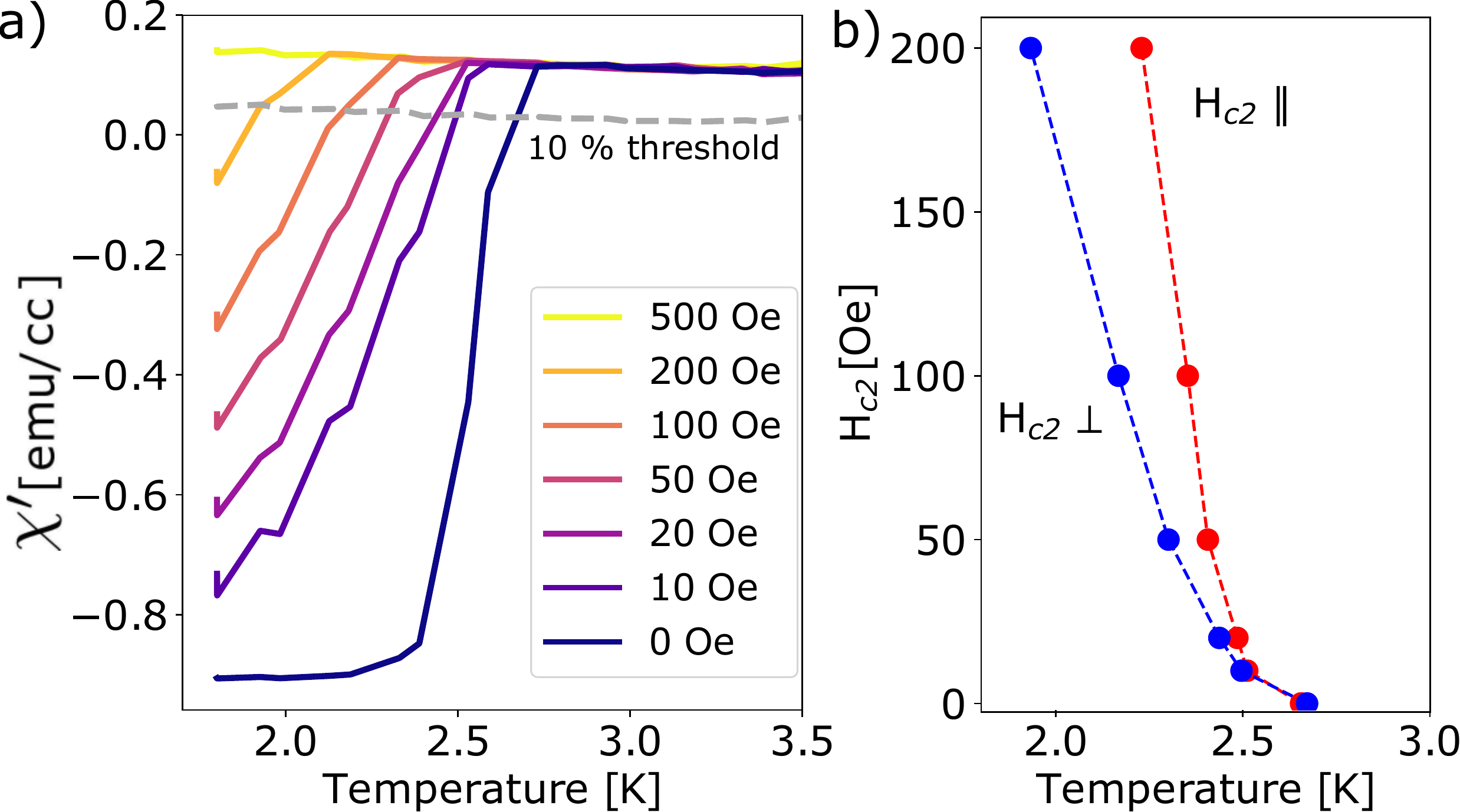}
\caption{AC susceptibility data for a 2.8\% intercalated sample. a) In-plane inductive component vs. temperature for various fields. The dotted line represents the 10\% transition threshold from which the $H_{c2}$ values are derived. b) $H_{c2}$ for both in-plane and perpendicular measurements.}\label{acsusc} 
\end{figure}

In-phase ($\chi'$) component of the AC susceptibility for a sample with 2.8\% Pd intercalation is shown in Fig.~\ref{acsusc} at various values of applied field.  
Here $T_c(H)$ was determined by the onset of the transition in the inductive term $\chi'$, which is approximated by the point at which the data reached $\sim$10\% of the total transition at zero field.  This criterion typically coincides with the zero-resistance transition and in general is free of complications due to sample shape and gross inhomogeneities. 
At zero magnetic-field, the peak in $\chi^{\prime\prime}$ is used, which yields very close transition values to the onset of $\chi^{\prime}$. As a summary, zero-field $T_c$ from the AC susceptibility peak are plotted in  Fig.~\ref{pd}, along with previous measured values from \cite{Straquadine2019}. 

Examination of the $H_{c2}$ data clearly show positive (upward) curvature near $T_c(0)$  for both parallel and perpendicular fields (also observed in the mid-point of the transition). 
Such a phenomenon is ubiquitous to layered compounds 
where it has been ascribed to the presence  of strong anisotropies of the Fermi surface \cite{Woollam1973,Wexler1976,Dalrymple1984}, possibly also a result of anisotropic electron-phonon interactions. (This positive curvature is opposite to dimensionality crossover  \cite{real_space}, and has been modeled with superconductivity that originates in the planes, and is coupled to the next layer through a proximitized intermediate layer \cite{Theodorakis1988}, a model that was shown to work well with the more anisotropic cuprate superconductors \cite{Theodorakis1989}.) While this
curvature complicates the analysis of $H_{c2}$, particularly the determination of the coherence length, it reinforces our previous observations that even below the secondary CDW transition, the Fermi surface remains anisotropic \cite{Brouet2008,fang2019disorder}. Another striking observation is the strong anisotropy of $H_{c2}$, especially past the initial positive curvature. Indeed, such strong anisotropy is expected and was previously observed in resistivity measurements of the  pristine compound where $\rho_b/\rho_{a-c}\sim 100$ (see e.g. \cite{Pfuner2010}). 

While positive curvature complicates the determination of the in-plane coherence length, we can still try to impose a very low-field slope while using the ``standard" WHH theory \cite{werthamer1966temperature}. Here $H_{c2\perp}(0)=0.69T_c \frac{dH_{c2\perp}}{dT}\Big|_{T=T_c}=200$ G, 
 which by examination of Fig.~\ref{acsusc}, is much too low. (200 G gives $\xi_0=\sqrt{\frac{\Phi_0}{2\pi H_{c2\perp}(0)}}\approx 1500$\AA.)  At the same time, we are sure that the normal state is almost completely restored in a perpendicular field of 500 G, corresponding to $\sim 950$\AA. In fact, STM data that will be discussed next may point to $H_{c2\perp}(0)\approx 210$ G for a  $T_c \approx 1.4$K, 2\% sample, thus, if scaled to a $T_c\approx 2.6$ K sample, yields $H_{c2\perp}(0)\approx 400$ G. In general, we find that these lengths are robust and were found to be consistent between samples measured with 3.3\%, 3.6\% and 4.1\% Pd intercalation as well.  
 
 With decreasing intercalation, $T_c$ starts to drop rather sharply around  $\sim$2\%, whereas for all intercalations above 2\%, the $T_c$s vary between 2-3 K (most likely due to sample variations). Thus we conclude that intercalation above 2\%, while further disrupting the CDW order, does not have a significant effect on the superconducting properties. These results constrain our determination of the in-plane coherence lengths, which allow us to analyze the STM data below. To estimate the penetration depth, we use a parallel field $H_{c1}\approx 5$ G that was measured for $\sim4.3 \%$ intercalation \cite{Straquadine2019}. With an anisotropy ratio $\sim 100$, determined from resistivity measurements (see e.g. \cite{Pfuner2010}), we estimate the angular averaged in-plane penetration depth as $\lambda_\perp\approx 2000$\AA, yielding a Ginzburg-Landau parameter of $\kappa=\lambda/\xi\approx 2-3$. This also yields an estimated $T_\theta \gtrsim170$K for the scale of phase ordering, thus, it is reasonable to expect that phase fluctuations play very little role in determining $T_c$.
With decreasing intercalation and $T_c$, $H_{c1}$ becomes difficult to determine. However, it is evident from the STM studies that the material remains type-II throughout where superconductivity is observed. 

\subsection{Pristine ErTe$_3$ and the emergence of superconductivity}
\begin{figure}[h]
\includegraphics[width=\columnwidth]{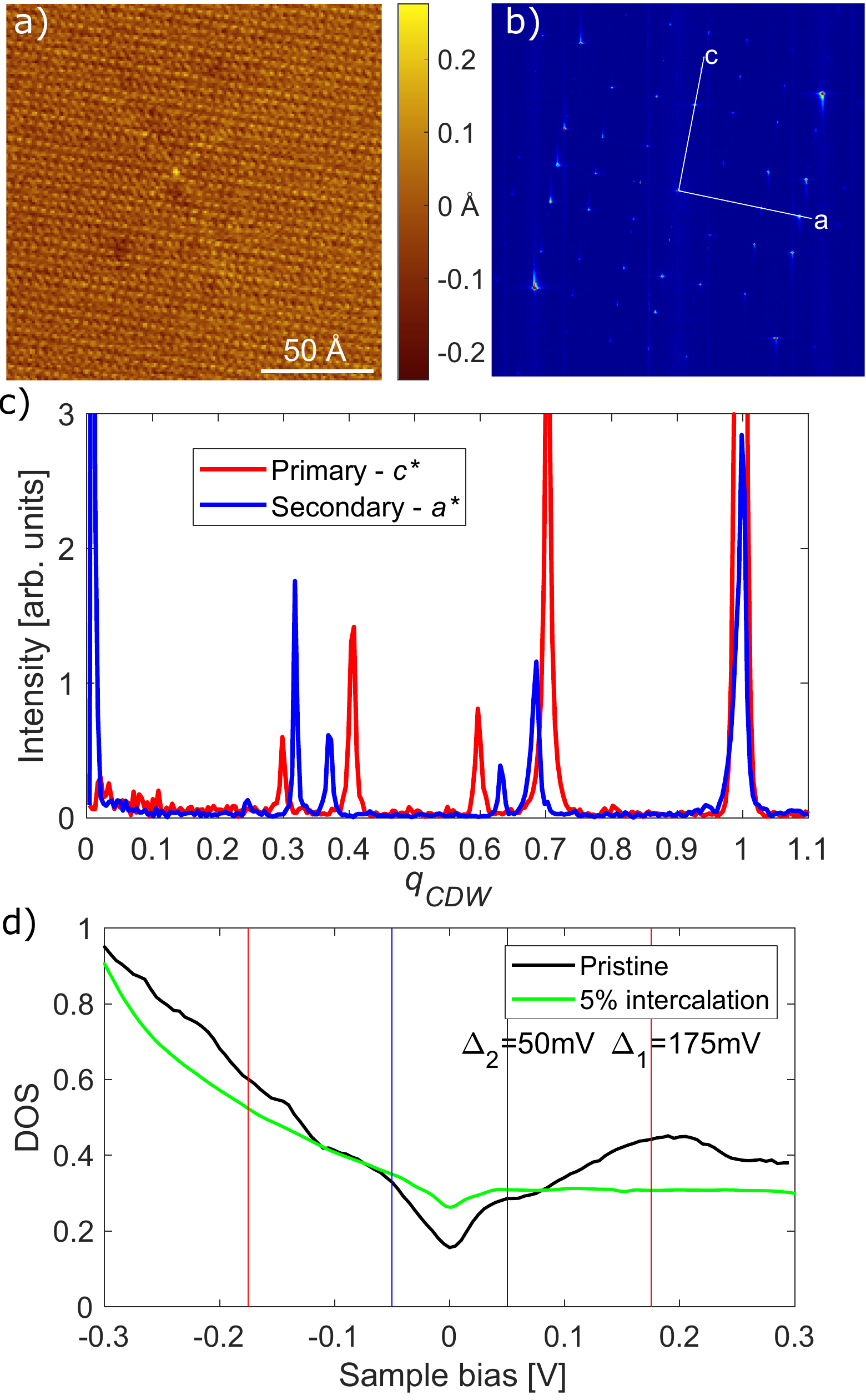}
\caption{a) Cropped topography of the pristine compound ErTe$_{3}$ at a temperature of 1.7 K.  b) Fourier transform showing the atomic lattice points, CDWs, and satellites.  c) Line cuts differentiate the two CDWs, labelled here by their wavevectors $q_{CDW1} = 0.70c^*$ and $q_{CDW2} =0.68a^*$. d) $dI/dV$ spectrum (both pristine and $x \sim$ 5\%) with lines indicating the two CDW gaps as determined by photoemission. }
\label{zerop} 
\end{figure}

To establish a baseline for the STM data at the scale of the superconducting gap, we first show results for the pristine (no intercalation) sample, in which superconductivity is not seen down to 400 mK in STM, nor 100 mK in heat capacity measurements.  Fig.~\ref{zerop}a contains a cropped image of the surface topography.  At this scale, the surface atomic lattice is visible, as well as the primary CDW corrugation in the $c$-axis.  The surface is generally free from defects, but there are occasionally two types of visible defects.  Both are likely to be sub-surface since the topographical height variations are sub-angstrom.  One is a missing-atom type defect represented by a shallow pit with no other features around it.  The other is a protrusion with faint streaks nearby in a slightly compressed $\times$ pattern.  (These are described in more detail in a previous work \cite{fang2019disorder} and represent quasiparticle scattering).  We identify the two perpendicular CDWs via Fourier transform (Fig.~\ref{zerop}b) and line cuts in Fourier space (Fig.~\ref{zerop}c).  Due to the small amounts of disorder, the CDW peaks are very sharp and well-defined.

To 
investigate the emergence of superconductivity in this compound, we first analyze the density of states at the energy scale of the CDW. A typical local $dI/dV$ spectrum is shown in Fig.~\ref{zerop}d, where within the wide range of bias voltage shown the superconducting gap is not visible. 
The 
vertical lines mark the photoemission-determined (maximal) gap energies for the primary (175mV) and secondary (50mV) CDW gap energies \cite{Moore2010}.  
There is a shoulder/depression in the DOS just below the two positive sample bias gap energies, which is expected for the decreased DOS inside a gap.  Since the Fermi surface is only partially gapped by the two CDWs \cite{Moore2010}, a finite density of states around zero-bias remains;  this may allow for the emergence of (not yet detected) superconductivity even in the pristine material. However, there is an increase of about 75\% in DOS near zero bias between the pristine and 5\% intercalated sample, which could account for the emergence of superconductivity with $T_c\approx 2.5$K. Since in BCS 
theory, the DOS appears in the exponential, small variations in the BCS coupling constant, $\lambda$,  may lead to strong variations in $T_c$. 
Taking an average of in-plane phonon frequency of $\sim 110$ cm$^{-1}$ \cite{Lazarevi2011}, we estimate for the 5\% sample $\lambda_{5\%}\approx 0.23$. If we assume that intercalation dependent changes in $\lambda$ arise exclusively from changes in the DOS, the suppressed DOS near zero bias for the pristine sample would yield  $\lambda_{0\%}\approx 0.13$, which in turn predicts $T_c\sim 110$ mK.  This is right at the  
low temperature limit of the range of $T$ over which this material
has been investigated.  Indeed, as we demonstrate below, in the 0.3 \% sample, where both CDW transitions are almost the same temperature as in the pristine sample, the superconducting $T_c$ is 0.76K.  

\subsection{Scanning Tunneling Microscopy and Spectroscopy of intercalated samples}

With intercalation of Pd atoms, the CDW becomes disordered and weakened, while superconductivity emerges at observable temperatures. However, with the underlying bi-directional CDW, and the disorder induced by the intercalation atoms, a first question that arises with respect to superconductivity is its uniformity. To check if there are spatial variations in the superconducting gap, we took spectroscopic scans which consist of a $dI/dV$ spectrum at each point within an area. For higher intercalations, we chose smaller areas to reflect the increasing density of dislocations, thus increased disorder.  Fig.~\ref{spectra} shows  histograms of all spectra over a particular area for three different doping levels, together with a s-wave fit to the data. The different effective temperature for each measurement is related to the electronic noise for that set of data and is discussed in detail in the Appendix.\\

\noindent  {\it 0.3\% intercalated sample}:

\begin{figure}[h]
\includegraphics[width=0.96\columnwidth]{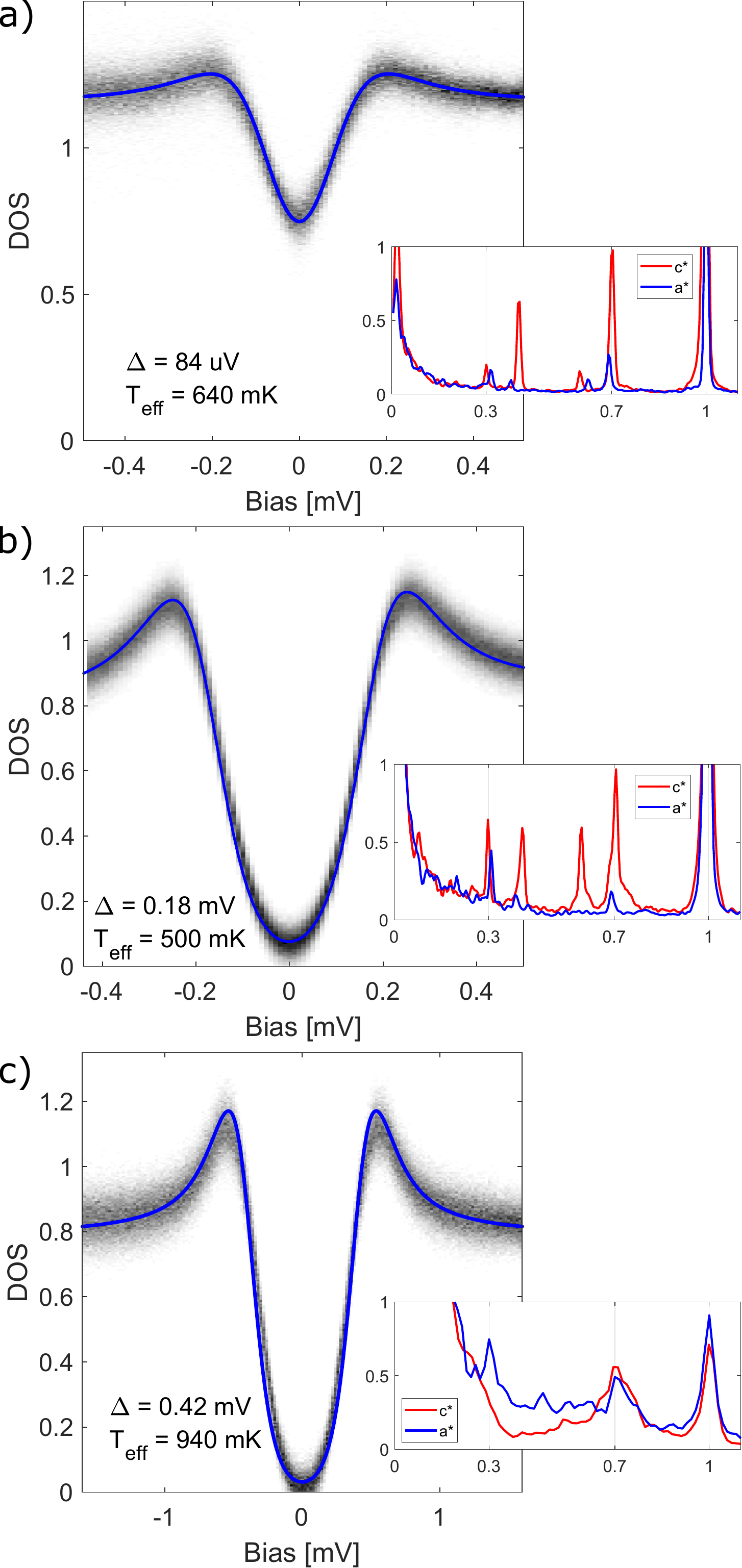}
\caption{Histogram of all spectra over a given area and the respective s-wave fit: a) (4000 \AA)$^2$ of 0.3\% intercalated sample; b) (300 \AA)$^2$ 2\% sample; c) (128 \AA)$^2$ 5\% sample. 
Insets show FFT linecuts along the primary ($c^*$) and secondary  ($a^*$) CDW directions, demonstrating the observation of short range CDW order with increasing intercalation (see ref.~\cite{fang2019disorder}).}
\label{spectra}
\end{figure}
We first discuss the 0.3\% sample, where superconductivity has set in but the $T_c$ is still below the saturation value at higher intercalations. At this level, no surface Pd atoms were visible. (Since Pd intercalates between the two Te sheets that cleave, most likely it is volatile and left the surface after our room-temperature cleave.)  As we have previously shown \cite{fang2019disorder}, despite the slightly lowered transition temperatures for the two CDWs, they are largely unaffected in terms of wave-vector.  In Fig.~\ref{spectra}a we show a histogram of all spectra over (4000 \AA)$^2$ and a s-wave fit with $\Delta_{Teff}$ = 84 $\mu$V and $T_{eff}$ = 640 mK.  The  ``fuzz'' of noise in the spectra histogram represents the amount of electronic measurement noise in each of the individual spectra, but otherwise the spectra are extremely uniform within this area.  Note that the DOS at zero bias is a significant fraction of the normal DOS, meaning that the measurement temperature is already near $T_c$.  To obtain the zero-temperature gap $\Delta_0$ and $T_c$, we assume the relation $e\Delta_0=1.76k_BT_c$ and compare this $dI/dV$ curve to a theoretically-predicted one given the $\Delta$ vs $T$ curve of other known s-wave superconductors to obtain $T_c$ = 0.76K (plotted in Fig. \ref{pd}) and $\Delta_0$ = 0.12 mV.\\

\noindent  {\it 2\% intercalated sample}:

With higher levels of intercalation, the superconducting $T_c$ rises while $T_{CDW}$ falls.  While from the anisotropic resistivity measurements in Fig.~\ref{pd}, the secondary CDW is expected to be fully suppressed,   we showed previously \cite{fang2019disorder} that in fact it still exists at low temperature, although with a higher dislocation density.  Thus the CDW peaks in the line cut of Fig.~\ref{spectra}b are slightly wider. Despite a significantly higher density of intercalants, the histogram of all spectra (over a smaller area this time, (300 \AA)$^2$) still show little spatial variation, with $\Delta_{Teff}$ = 0.18 mV and $T_{eff}$ = 500 mK.   (We also took individual spectra outside of the scan area to confirm uniformity at longer distances.) Since the DOS at zero-bias is very low, we are now only at a small fraction of $T_c$.  Thus $\Delta_0\approx 0.21$ meV, and the predicted transition temperature is $T_c \approx1.4$K.  Note that this is significantly lower than $\approx$2.5K measured at a similar intercalation in the AC susceptibility measurement.  We do not believe this to be a surface effect from STM measurement, nor differences in the condition at the cleaved surface, since the coherence length is large and the bulk superconductivity should proximitze to the surface.  Instead, it is likely a consequence of  $T_c$ varying rapidly around 2\%, and this particular sample being slightly below the threshold.\\

\noindent  {\it 5\% intercalated sample}:

At 5\% intercalation levels, our previous STM study showed that the sample 
exhibits a ``vestigal nematic'' state \cite{fang2019disorder}, 
while X-ray and electron diffraction reveal that the sharp superlattice peaks observed for low Pd concentrations are now replaced by broad and diffuse streaks spanning between the original CDW points, indicating short range CDW correlations that are consistent with the $q$-dependent susceptibility \cite{Straquadine2019}.
 
Here, the superconducting T$_c$ is at its full value of $\approx$ 2.5 K as measured by AC susceptibility.   At this level of intercalation, the sub-\AA~ height variations, which are likely from the Pd in the sub-surface layers, is much more prevalent. While large area scans have
proven to be difficult to perform because of Pd adatoms on the surface (roughly once every $\approx$ 300 \AA), we were able to perform limited spectroscopic scans, both at high and low energies.    Fig.~\ref{spectra}c is a histogram of all the $dI/dV$ spectra over (128 \AA)$^2$ and a fit to a s-wave BCS spectrum with $\Delta_{Teff}$ = 0.42 mV and $T_{eff}$ = 940 mK. Using a value of $\Delta_0$ of 0.44 mV, we expect a $T_c$ of 2.9K, which is slightly above those from AC susceptibility.  This difference could be because of disorder-induced increased broadening of $\chi^{\prime\prime}$'s dissipation peak, that was used to determine $T_c$. Superconductivity, at least in terms of gap size and other spectral features, is very uniform over this area despite the large amount of disorder.  We also took individual spectra over the microscope's scan range of $\approx$ (5000 \AA)$^2$ and did not notice any variations. 

In terms of CDW features, Fig.~\ref{zerop}c shows the average high bias spectrum also taken over (128 \AA)$^2$.  While the individual spectra do show spatial variations, they are mainly correlated with topographic features (and not presented here).  The CDW gaps are largely wiped out, aside from a tiny dip at $\approx$ 150 mV and a slight depression in the DOS at zero bias.  The curve is largely feature-less aside from a rising DOS at large negative bias.  Similarly, the CDW peaks are very broad in the FFT linecuts of Fig.~\ref{spectra}c inset.\\
 
\noindent  {\it Clean vs. Dirty limits}:

As noted in the susceptibility section, the positive curvature of $H_{c2}$ is likely a result of the layered nature of ErTe$_3$, and the in-plane Fermi surface anisotropy. While in-plane anisotropy of the Fermi velocity has been inferred for all RTe$_3$ \cite{Sinchenko2014}, the local spectroscopy averages the in-plane directions. An averaged Fermi velocity over the Fermi surface seems to be invariant for all rare-earth tellurides: RTe$_3$ at $v_F = 10 \pm 1$ eV-\AA~from photoemission \cite{Brouet2008}. Using this value we can estimate the BCS coherence length for this material $\xi_{BCS}=\hbar v_F/\pi\Delta_0$. Assuming that the in-plane Fermi velocity is very weakly dependent on intercalation, the coherence lengths that we extract are $2.65~\mu$m, $1.5~\mu$m, and $0.72~\mu$m for the 0.3\%, 2\% and 5\% intercalation respectively. The Ginzburg-Landau (GL) coherence length, which takes into account scattering, can be inferred from the upper critical field as determined in the STM measurements. Starting from the 2\% data, we observe that the gap structure disappears at $\sim 210$ G, yielding $\xi_0(2\%)\approx 1250$\AA, which is much smaller than the BCS coherence length, and thus implies that the materials must be in the ``dirty limit,'' i.e. $\xi_0 \approx \sqrt{\xi_{BCS} \ell}\ll \xi_{BCS}$ where $\ell$ is the elastic mean-free path. In fact, this conclusion holds for all intercalation levels above $\sim2\%$. At lower doping $T_c$ tends rapidly towards zero, thus the BCS coherence length is fast increasing, which probably maintains the dirty limit throughout the intercalation range.\\

\noindent  {\it Vortex-core and coherence lengths}:

\begin{figure}[h]
\includegraphics[width=1.0\columnwidth]{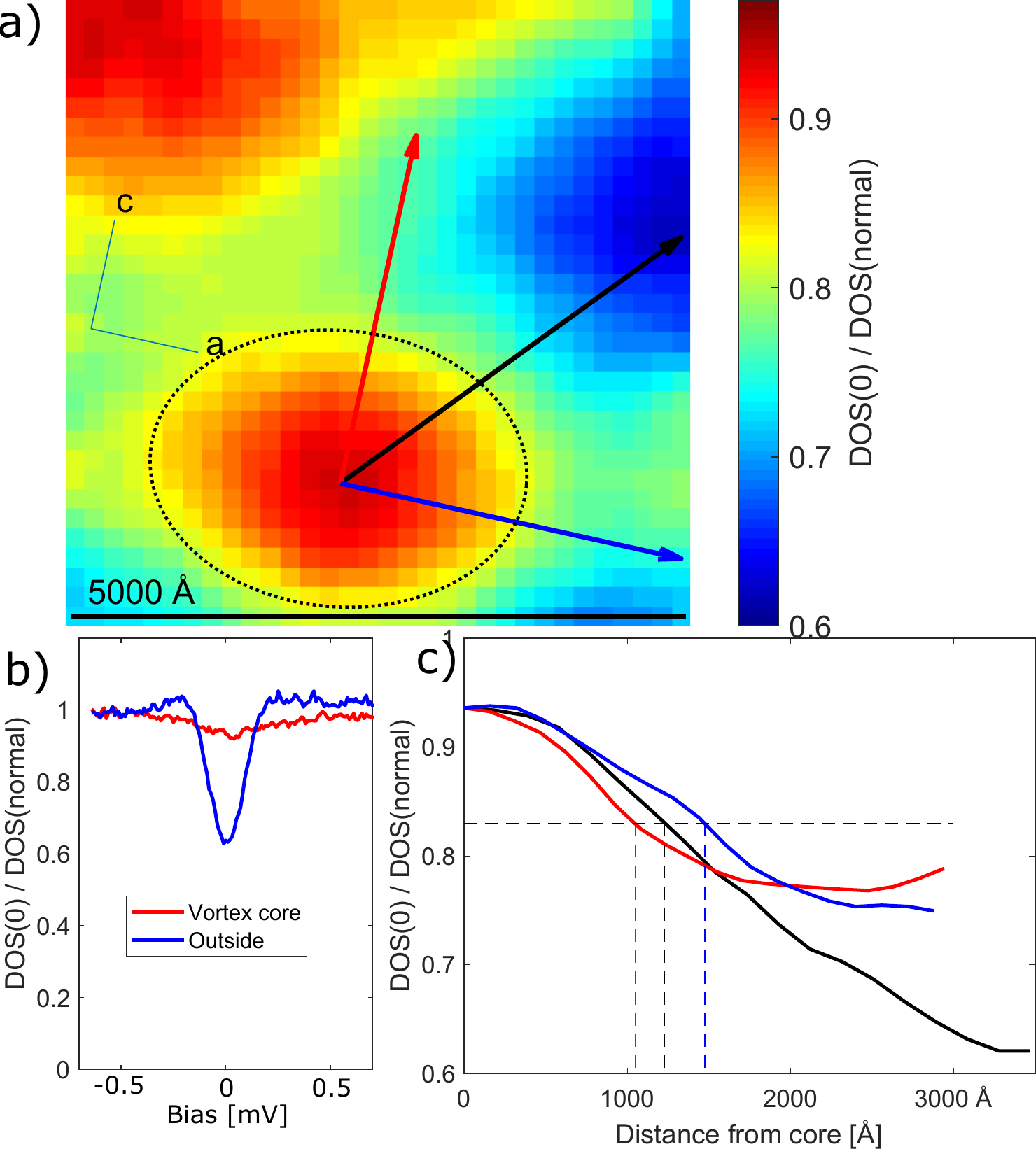}
\caption{2\% intercalated sample in magnetic field. a) an elongated vortex recorded at 110 G and its environment b) Example spectra inside and outside of a vortex c) Line cut from vortex core to outside, along the crystal's principal axes and at $\sim 45^o$ towards a ``pit'' of minimal conductance.  Line cuts follow the arrows in (a), and the dashed horizontal line corresponds to the same level as the rim of the vortex core as marked by an ellipse.}
\label{twopercentvortex}
\end{figure}
Another way to determine the coherence length and its anisotropy is via the vortex core size and shape.  We note that upon going above and then below $T_c$ again (for a $^3$He condensation cycle), the vortices do not show up in the same location, meaning that vortex pinning is weak.  This concurs with the small magnetic hysteresis loops observed in previous measurements \cite{Straquadine2019}. This, plus the fact that the vortex spacing is large due to a low $H_{c2}$, in turn prevents us from large-scale vortex imaging to determine the symmetry of the vortex lattice. Thus, focusing on individual, ``zoomed-in'' vortices, Fig.~\ref{twopercentvortex}a shows a (5000 \AA)$^2$ area conductance map (zero-bias DOS normalized by the normal state DOS at energies outside the gap) at a magnetic field of 110 G, where two vortices are clearly visible. Two sample spectra inside and outside the vortex shows the almost normal state DOS at the center of the core, with a depressed ($\sim$ 60\% of normal) DOS outside the core (Fig.~\ref{twopercentvortex}b), as expected at a field which is about half of $H_{c2}$.

The elongated shape of the vortex core points to an anisotropic coherence length. To estimate that anisotropy we show in Fig.~\ref{twopercentvortex}c linecuts along the two principal axes: $c$-axis which is the primary CDW direction, and $a$-axis which is the secondary CDW direction. The scales extracted from the width at the rim of the vortex (compared to the surrounding background) suggest $\xi_a\approx1500$\AA\ and $\xi_c\approx 1000$\AA. The average of $\approx 1250$\AA~that we determined from $H_{c2\perp}(0)$ fits very well within this range of scales.

\section{Discussion}

The Pd intercalated rare earth tellurides present a particularly interesting perspective on the interplay between CDW, orientational (nematic) order, superconductivity and disorder. At room temperature the crystal structure of the pristine material consists of alternating ErTe slabs with bilayers of square Te nets.  It is approximately tetragonal, although 
the presence of a glide plane in the stacking of these layers creates a 0.05\% \cite{Ru2008} difference between the in-plane $a$- and $c$-axis lattice parameters at room temperature,  and this in turn biases the primary CDW transition to order along the c axis 
 at $T_{CDW1}\approx 270$K. At that point the (almost) 4-fold symmetry is broken with noticeable Fermi surface effects. A secondary CDW transition occurs at $T_{CDW2}\approx 170$K, 
with additional changes to the Fermi surface \cite{Moore2010}. Despite a reasonably well-nested Fermi surface, CDW formation in these materials has been attributed to a strongly $q$-dependent electron-phonon coupling \cite{Maschek2015} 
with a focusing effect associated with corners of the diamond-like Fermi surface sheets where the CDW gap is largest \cite{Lavagnini2010}. Intercalated Pd atoms reside in the van-der-Waals gap between the two Te planes, and thus slightly affect the $b$-axis separation (as much as $\sim 0.28\%$ at $5\%$ intercalation) \cite{doping}
 
While intercalation disrupts the bi-diectional CDW order, vestiges of the two phase transitions have been observed even up to 5\% intercalation \cite{Straquadine2019,fang2019disorder}. 
Superconductivity 
with transition temperatures above 100mK 
emerges at relatively small $x$ and rapidly reaches a $T_c\approx 2.5$K. This evolution is very similar to the behavior of the pristine material under hydrostatic pressure, where  as the two CDW states are suppressed with pressure, superconductivity appears \cite{hamlin2009pressure,zocco2015pressure} at similar temperatures as with intercalation.  While very different in nature, the similarity in the result of these two behaviors is rather striking. 

The common wisdom of superconductivity in a CDW system is that the two compete ferociously, and thus when CDW is suppressed, superconductivity emerges (see e.g. \cite{bilbro1976theoretical}). Moreover, the explanation of this is tied to the accepted view that both orders are driven, energetically, by the gapping of states at the Fermi energy;  states that are gapped by one of these orders are not accessible to the other.  However, the data  we have presented shows no direct competition between CDW order and superconductivity,  as they both coexist over most of the intercalation range, with uniform superconductivity over length scales that exceed the superconducting coherence length. Moreover, the local superconducting gap is insensitive to defects in either of the CDW orders. At the same time, the elongated vortices (Fig.~\ref{twopercentvortex}) clearly show the effect  on the superconducting state of the Fermi surface distortions  that reflect the vestigial nematic order that remains, even when the long-range CDW order is disrupted.  (The state is nematic in the sense that the correlations associated with CDW1 along the $c$ axis remain substantially stronger than those of CDW2 along the $a$ axis.)

We conclude with some speculative comments about the implications of our findings for the microscopic theory of superconducting and CDW orders.  The fact that the zero energy DOS is suppressed from its normal state value by approximately 50\% in the pristine material is consistent with the conventional understanding that the opening of a gap on a portion of the Fermi surface is essential to the mechanism of CDW formation.  However, the fact that local CDW order remains pronounced over the entire range of $x$, while the partial gapping of the DOS at the Fermi energy is all but eliminated at higher $x$, suggests that the opening of a CDW gap should be viewed as a  {\em consequence} of CDW order, rather than the {\em cause} \cite{analogy}.
Correspondingly, the effectively gapless CDW state that occurs at non-zero $x$ does not, in fact, seem to compete with superconductivity to any significant extent. 
Such a perspective might be consistent with earlier commentaries which have noted that Fermi surface nesting alone is unable to drive CDW order in this (and other) quasi-2D materials \cite{Johannes2008}. Indeed, intuition drawn from a chemical bonding perspective for materials that have similar Te networks implies a tendency towards local bond order in the form of extended oligomers, since the Te-Te bond-length in the RTe$_{3}$ structure above $T_{CDW}$ is considerably longer than the usual value \cite{Patschke2002,Kim2006}.

It has been suggested \cite{Kivelson1998}, in the context of the cuprates, that another form of competition between CDW order and superconductivity concerns the superfluid density;  at an intuitive level, electrons that form the CDW condensate cannot at the same time contribute to the superfluid stiffness.  However, while the superfluid stiffness plays an essential role in determining $T_c$ in the cuprates (through its role in determining the extent of phase fluctuations), phase fluctuations likely play little role in the present materials.  In particular, an estimate\cite{vicandme1995} of the temperature at which phase-fluctuations alone would destroy superconducting order yields a value $T_{\theta} \approx 170$ $\gg T_c$.  

Finally, this line of reasoning implies the prediction that the pristine material should have a $T_c$ which may be only slightly lower than the lowest temperatures explored to date.  Specifically, as discussed above, if we assume that intercalation dependent changes in the dimensionless coupling $\lambda$ are determined solely by changes in the Fermi energy DOS, we would predict a $T_c \approx 110$mK.   The applicability of `Anderson's theorem,' \cite{Anderson1959} implies insensitivity of the superconducting state to other aspects of the disorder.  The momentum dependence of the electron-phonon coupling somewhat complicates this analysis, as it implies that some portions of the Fermi surface could be more suitable for superconductivity than others, but we think it is likely that this does not change the proposed analysis qualitativelty.

\section*{Acknowledgments}
  This work was supported by  the U.~S.~Department of Energy (DOE) Office of Basic Energy Science, Division of Materials Science and Engineering at Stanford under contract No.~DE-AC02-76SF00515. JS is supported by an ABB Stanford Graduate Fellowship. AGS was supported in part by an NSF Graduate Research Fellowship (grant number DGE-1656518). Various parts of the STM system were constructed with support from the Army Research Office (ARO), grant No.~W911NF-12-1-0537 and by the Gordon and Betty Moore Foundation through Emergent Phenomena in Quantum Systems (EPiQS) Initiative Grant GBMF4529. 

\clearpage

\setcounter{figure}{0}

\renewcommand{\thefigure}{A\arabic{figure}}

\section*{Appendix: Analysis of the superconducting gap spectroscopy}
STM is a powerful tool for the determination of superconducting properties of superconductors. In particular, spectroscopy mode allow for studies of gap variations on the atomic scale, searching for uniformity over macroscopic areas, as well as broadening due to intrinsic effects in the material, such as scattering and non-equilibrium effects. The relevant measure is the tunneling conductance, which is extracted from the derivative $dI/dV$ of the tunneling current $I(V)$: 

\be
I(V,T)\propto \int_{-\infty}^\infty \nu(E)\nu_{tip}(E-V)\left[f(E,T)-f(E-V,T)\right]dE
\ee
Here $f(E,T)$ is the Fermi function evaluated at an energy $E$ and temperature $T$, and $\nu(E)$ is the density of states of the sample at the energy $E$. For good conductors, where $V\ll E_F$ for either the sample or the tip, and a tip density of states that is featureless on the scale of $V$, the derivative $dI/dV$ simplifies to:
\be
\frac{dI}{dV} \propto \int_{-\infty}^\infty \nu(E)\frac{df(E-V,T)}{dV}dE  
\ee
which is proportional to the sample's density of states convolved with the derivative of the finite-temperature Fermi function - a thermally broadened peak.  

For a BCS superconductor we expect the sample density of states
\be
\nu(E)=\Re\Bigg\{\frac{|E|}{\sqrt{(E^2-\Delta^2)}}\Bigg\}\;\;\; \nu=0\; \mathrm{for}\; |E|<\Delta
\ee
However, Dynes {\it et al.} \cite{Dynes1978} noticed broadened spectra in tunneling studies of a strongly coupled superconductor,  and proposed to add a lifetime-broadened energy gap edge parameter into the density of states by adding an imaginary part to the energy:
\be
\nu(E,\Gamma)=\Re\Bigg\{\frac{E-i\Gamma}{\sqrt{(E-i\Gamma)^2-\Delta^2)}}\Bigg\}
\ee

\begin{figure}
\includegraphics[width=\columnwidth]{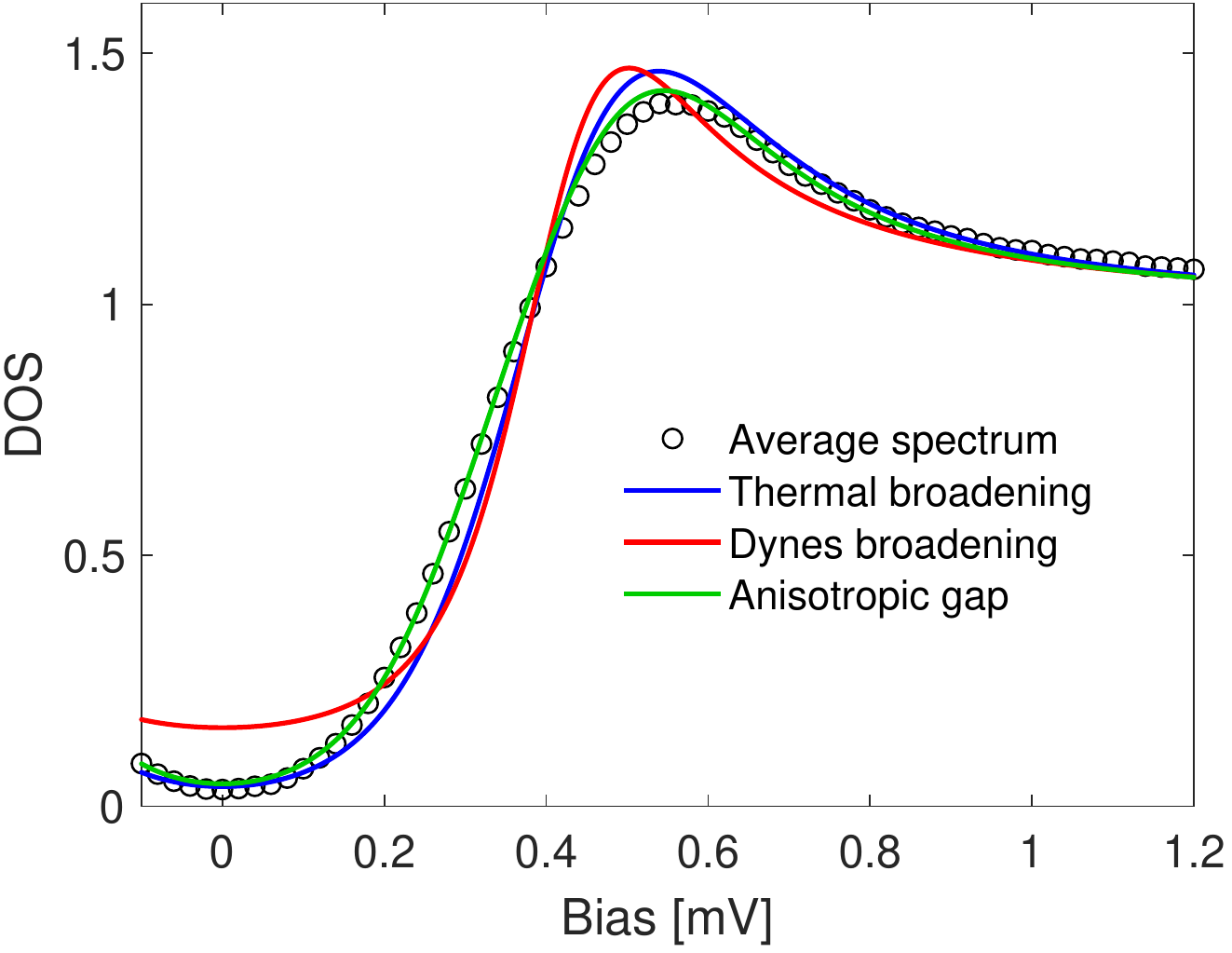}
\caption{Spatially averaged spectrum for a 5\% intercalation sample.  Fit using primarily thermal broadening (blue) has parameters of $\Delta$ = 0.42 mV, $T_{eff}$ = 940 mK, $\Gamma$ = 2 uV.  Dynes broadening (red) uses $\Delta$ = 0.42 mV, $T_{eff}$ = 400 mK, and $\Gamma$ = 65 uV.  Anisotropic gap fit uses $\Delta$ = 0.40-0.50 mV, $T_{eff}$ = 840 mK, $\Gamma$ = 2 uV.}
\label{broad} 
\end{figure}

In our fits to the spectra, we considered both of these effects for spectral broadening over what is expected from the base measurement temperatures of $\leq$400mK.  What we found was that adding Dynes broadening led to an unrealistically high zero-bias DOS inside the gap, whereas adding additional thermal broadening via a higher ``effective temperature'' made for a better fit.  An example is provided in Fig.~\ref{broad}.

For the Dynes broadening case, 400 mK of thermal broadening is added to reflect the measurement temperature.  In the case of the thermal fit, a very small amount of Dynes broadening is added (only for numerical reasons) so that the zero-temperature superconducting DOS does not diverge at the coherence peaks.  In many measurement techniques at low temperature (i.e. below 4K), a primary cause of a higher``effective temperature'' is heating from Radio Frequency (RF) noise.  Thus RF filters are usually applied to all the wires leading into the experiment. In our measurements on the 0.3\% and 2\% samples, the effective temperature from the fit is only a little higher than our microscope with the RF-filters (tested at $\approx$ 500 mK).  Our measurements on the 5\% sample were done while there were RF filters on only some of the wires, hence the higher effective temperature. 

Note that even in the thermal fit, the real data has shorter and blunter coherence peaks and also has some additional mid-gap DOS.  One possibility is a slightly anisotropic gap structure (e.g. $\Delta$ is slightly different in the $a$ and $c$ axis).  By having the gap vary between 0.40-0.50 mV, an even better fit is achieved via the green curve, while the resultant thermal broadening term is also slightly lower and more in line with our previous tests ($\approx$ 800 mK).  Note that we rule out the possibility of real-space gap variations, since the histogram of Fig.~\ref{spectra}c in the main body of the paper shows very little mid-gap variation.  Other possibilities, such as a two-gap superconductor, would allow for even more fitting parameters, but are not considered here.

\bibliography{pdxerte3_sc}

\begin{thebibliography}{35}
\bibitem [{ana()}]{analogy}%
  \BibitemOpen
  \href@noop {} {}\bibinfo {note} {We note that a possibly interesting analogy
  can be drawn here with the CDW correlations seen in the cuprates, which while
  apparently ubiquitous in the hole-doped cuprates, and directly visible as
  local order in STM studies, does not imprint itself in any obvious way on the
  quasi-particle spectrum as measured in ARPES.}\BibitemShut {Stop}%
\end{thebibliography}

\end{document}